\documentstyle[twocolumn,aps,epsfig]{revtex}

\begin{document}
\twocolumn[
\hsize\textwidth\columnwidth\hsize\csname@twocolumnfalse\endcsname
\draft

\preprint{conmatt}
\title{Impurity-Induced Bound Excitations on the Surface of Bi$_{2}$Sr$_{2}$CaCu$_{2}$O$_{8}$}
\author{Ali Yazdani$^{1,2,*}$, C. M. Howald$^3$, C. P. Lutz$^2$, A. Kapitulnik$^3$, and D. M. Eigler$^2$}
\address{$^1$Department of Physics, University of Illinois at Urbana-Champaign, 1110 West Green Street, Urbana, Illinois 61801 \\
$^2$IBM Research Division, Almaden Research Center, 650 Harry Road, San Jose, California 94005 \\
$^3$Department of Applied Physics, Stanford University, Stanford, California 94305 \\}
\date{{\it To appear in Physical Review Letters}}
\maketitle
\begin{abstract}
We have probed the effects of atomic-scale impurities on superconductivity in Bi$_{2}$Sr$_{2}$CaCu$_{2}$O$_{8}$ by performing low-temperature tunneling spectroscopy measurements with a scanning tunneling microscope. Our results show that non-magnetic defect structures at the surface create localized low-energy excitations in their immediate vicinity.  The impurity-induced excitations occur over a range of energies including the middle of the superconducting gap, at the Fermi level. Such a zero bias state is a predicted feature for strong non-magnetic scattering in a d-wave superconductor.

\end{abstract}
]
\narrowtext

There is now a great deal of evidence that the superconducting state in a number of high-T$_{c}$ superconductors has a dominant d$_{x^{2}-y^{2}}$ symmetry \cite{Vanharlingen}. One key evidence is that non-magnetic impurities, which only affect conventional s-wave superconductors weakly, can act as strong pair-breakers in the high-T$_{c}$ superconductors. \cite{Annette,Lee} However, to date, there are no atomic scale studies of
superconductivity in the immediate vicinity of individual impurities in a high-T$_{c}$ superconductor. Such experiments are motivated by
recent theoretical studies of the local response of a d-wave superconductor to
individual impurities. These theories \cite{Byers,Balatsky1} predict local signatures of
d-wave pairing and emphasize the importance of local variations in the
electronic properties of a d-wave superconductor. \cite{Balatsky2,Franz1,Walker1}
There is also now growing theoretical and experimental evidence
that quasi-particle scattering from surfaces and twin boundaries of d-wave
superconductors gives rise to effects that have no analog in conventional s-wave
superconductors. \cite{Buchholtz,Hu,Green,Sauls,Walker2,Sigrist}

The scanning tunneling microscope (STM) offers a direct method to examine the spatial variation of the electronic properties in the superconducting state near defect structures.\cite{Yazdani} In this paper, we report on STM measurements of the local density of states near regions of a d-wave superconductor that have been perturbed by the presence of atomic-scale impurities. Our results show that non-magnetic impurities induce low-energy excitations in a d-wave superconductor, which are localized on length scales comparable to the superconducting coherence length.  Such impurity-bound excitations for non-magnetic impurities is a predicted signature of a d-wave superconductor \cite{Balatsky1} and is in stark contrast to behavior of conventional s-wave superconductors, in which to create similar effects, the impurities need to be magnetic.\cite{Yazdani} More specifically, we observe that the impurity-induced excitations for a d-wave superconductor can occur as a pronounced and narrow resonance at energies close to the Fermi level, E$_{F}$. Such a zero bias feature has been predicted to occur for impurity scattering in a d-wave superconductors in the unitarity limit \cite{Balatsky1}
or for strong impurities when considering the order parameter suppression.\cite{Shnirman} From a different perspective, a zero bias resonance has also been predicted and observed for Andreev scattering from interfaces at which quasi-particles experience a sign change of the d-wave order parameter and form a surface bound state at E$_{F}$.\cite{Hu,Green,Wei,Alff}
The similarity suggests that multiple Andreev scattering similar to that proposed at interfaces \cite{Hu} is at work in the bulk at strongly scattering impurities.

\begin{figure} 
\begin{center}
\epsfig{file=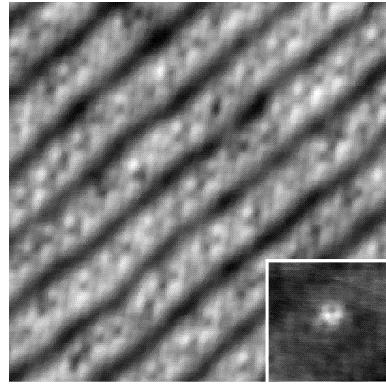,width=2.00in}
\caption{Main: Constant current STM topograph (200\AA\ $\times$ 200\AA) of the Bi$_{2}$Sr$_{2}$CaCu$_{2}$O$_{8}$ surface; $R_{J}=$ 250 M$\Omega$, $V=$ 0.25 eV. Inset: A topograph of a native surface defect (50\AA\ $\times$ 50\AA); $R=20M\Omega$, $V=$ 0.1V. The contrast (black to white) in the main image and the inset correspond to an apparent corrugation of 0.8\AA\ and 0.3\AA\ respectively.}
\label{autonum}
\end{center}
\end{figure}

We performed our experiments using an ultrahigh vacuum (UHV) STM which operates at low temperatures. The Bi$_{2}$Sr$_{2}$CaCu$_{2}$O$_{8}$ single-crystal samples were grown using directional solidification technique and re-oxygenated prior to the STM measurements. The experiments reported here were carried out on overdoped single-crystal samples with a superconducting transition temperature at 74K and a transition width of 3K, as characterized by magnetometry measurements. Samples were introduced into the UHV chamber at room temperature and mechanically cleaved prior to STM measurements performed at low temperatures (T=5K.)  We used a polycrystalline Au wire as our tip; however, the chemical identity of the last atom on the tip is unknown. The local quasi-particle density of states (LDOS) of the Bi$_{2}$Sr$_{2}$CaCu$_{2}$O$_{8}$ surface was obtained from measurements of the differential conductance $dI/dV$ (where I is the current) of the STM junction versus sample bias voltage V (with respect to the tip) performed under open feedback loop conditions. 
\begin{figure} 
\begin{center}
\epsfig{file=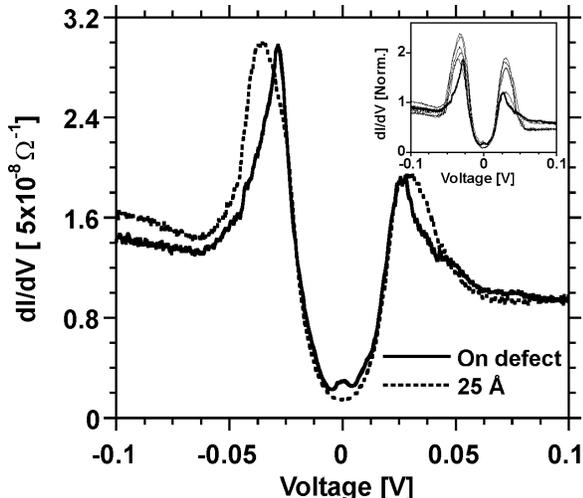,width=3.0in}
\caption{Main: Tunneling spectroscopy near the intrinsic surface defect shown in inset of Figure 1. Inset: The normalized spectra taken on different locations on two different samples; the data in the inset is normalized by the values of $1/R_{j}$ at -0.1V. The voltage refers to sample bias.}
\label{autonum}
\end{center}
\end{figure}
Figure 1 shows a constant current STM topograph of the cleaved Bi$_{2}$Sr$_{2}$CaCu$_{2}$O$_{8}$ surface. The weakest bond in the crystal is between the two adjacent BiO layers; therefore it is most likely that the top-most atomic layer of a cleaved sample is BiO.\cite{Kirk} The image in Figure 1 shows a long length scale modulation (period of 27\AA) which has been previously observed with the STM and has been associated with the relaxation of Bi atoms in BiO layer.\cite{Kirk}
Imaging the Bi$_{2}$Sr$_{2}$CaCu$_{2}$O$_{8}$ surface in several regions showed that there are intrinsic defects on the surface. The intrinsic impurities are difficult to identify, except in some cases, such as that shown in the inset of Figure 1, where defects appeared as protrusions in STM topographs.

The intrinsic defects are of fundamental interest, since they are considered crucial to understanding the deviations of low-temperature properties of high-T$_c$ superconductors from those expected for a clean d-wave superconductor.\cite{Annette,Lee,Balatsky2,Franz1} Our STM measurements near intrinsic surface defects show that such defects indeed modify the LDOS of Bi$_{2}$Sr$_{2}$CaCu$_{2}$O$_{8}$ surface. For example, Figure 2 shows the measurements of LDOS with the tip ``on" and ``off" the intrinsic surface defect shown in the inset of Figure 1. There is a clear enhancement of the LDOS at low-energies at this impurity site, as compared to that measured at lateral distance of about a coherence length away. The change in $dI/dV$ near $E_F$ corresponds to about 20\% of that measured at voltages above the superconducting gap. However, we found other intrinsic surface defects which did not modify the low-energy LDOS as much, but only affected the quasi-particle peaks or the LDOS background measured at 
higher voltages. Overall, the STM spectra over many regions of the ``bare" surface, away from any surface defects, were comparable to those previously reported in the overdoped regime.\cite{Renner} In the inset of figure 2 we show a collection of spectra from different regions of two samples with similar doping level. The values of the maximum gap ($2\Delta_{max}/k_b T_c\sim$ 9), the asymmetric background, and other features of our spectra such as the width of the quasi-particle peaks are similar to those previously reported by Renner {\it et al.} for a sample with a similar doping level and T$_c$. \cite{Renner} 

The intrinsic surface defects that do modify the LDOS at low-energies can be considered as evidence for impurity-induced excitations due to non-magnetic scattering in a d-wave superconductor. However, the magnetic state of such defects is unknown, hence to illustrate the main point of this work, we focus the rest of this paper on ostensibly non-magnetic defects created with the Au STM tip. Such defects were deposited onto the surface by bringing the Au STM tip close enough to the surface to cause transfer of atoms from the tip to the sample. We monitored the current through the STM junction during this procedure and retracted the tip once the current began to saturate signaling the formation of a small contact. In the regime of close contact, it has been shown that an STM junction resistance, $R_{J}\sim 25 k \Omega$, is indicative of formation of a single metal atom contact between the tip and the surface.\cite{Contact}  The STM topographs of the Bi$_{2}$Sr$_{2}$CaCu$_{2}$O$_{8}$ surface measured after making such contacts revealed the deposition of atoms from the tip onto the sample, similar to that observed when Au tips contact other conducting surfaces. An example of tip-deposited defects is shown in the topograph in Figure 3A, which shows a 36\AA\ square area of the surface with two defects appearing as protrusions with an apparent height of 0.8\AA.

The LDOS of the Bi$_{2}$Sr$_{2}$CaCu$_{2}$O$_{8}$ surface is dramatically modified by defects deposited from the Au STM tip. For the defect in the center of figure 3A these modifications are demonstrated by the data in figure 4A. This figure shows measurements of the LDOS with the tip over the defect structure (centered at maximum height) and that at lateral distances away (in the direction indicated in Figure 3A). Comparing these spectra shows that the defect induces low-energy excitations, within the superconducting gap, in its immediate vicinity. At the impurity site, the value of $dI/dV$ measured close to $E_F$ is about five times larger than that measured at voltages above the superconducting gap. A closer examination of data in Figure 2A shows that the impurity-induced resonance is made of two asymmetric peaks--one above and one below $E_F$ (within 1meV). It is also interesting to note that this asymmetric behavior is similar to that of the quasi-particle peaks measured with tip over a region far away from the impurity. 

\begin{figure} 
\begin{center}
\epsfig{file=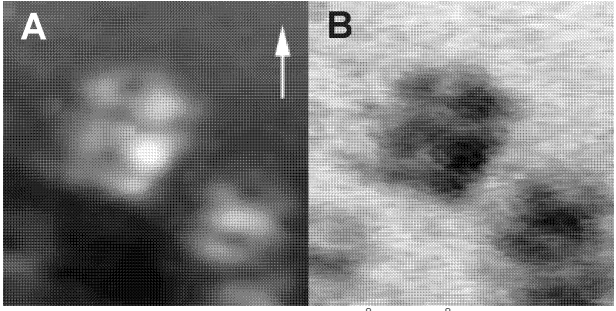,width=3.20in,height=1.6in}
\caption{(A) STM topograph (36\AA\ $\times$ 36\AA) of atomic-scale defects deposited from the tip; $R_{J}=$ 160 M$\Omega$, $V=$ 32 meV. Black to white corresponds 0.8\AA. These defects appear as protrusions also at higher V's.(B) Simultaneously acquired $dI/dV$ map. Black to white corresponds to a 50\% change in the signal. The areas where $dI/dV$ is reduced (dark) corresponds to the position of the bound state.[22]}
\label{autonum}
\end{center}
\end{figure}
\begin{figure}
\begin{center}
\epsfig{file=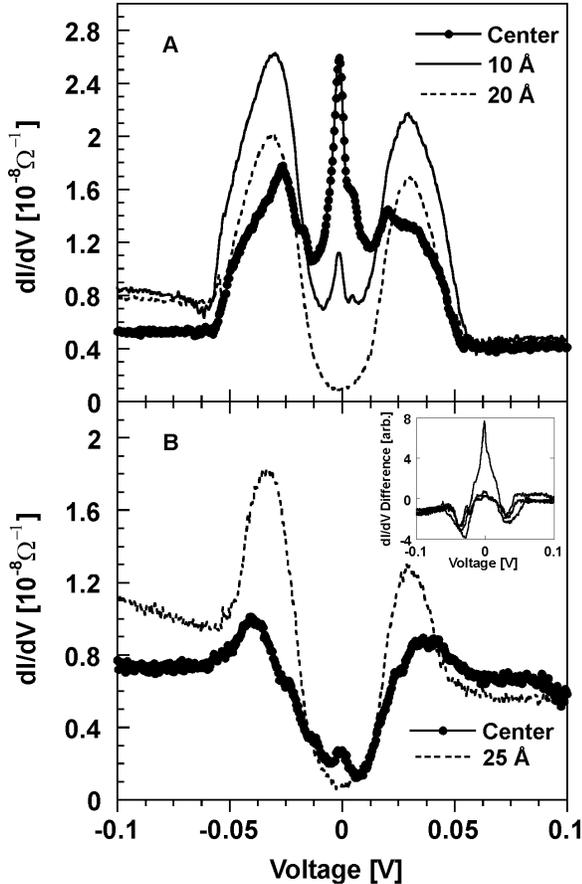,width=3.0in}
\caption{(A) Tunneling conductance near the defect shown in Fig. 3A; the distances indicated are those lateral from the center of defect structure along the direction indicated in Figure 3A. (B) Spectra taken near another tip-induced defect. The inset shows differences in the spectrum with the tip ``on'' and ``off'' different defect structures.}
\label{autonum}
\end{center}
\end{figure}
\vspace{1in}

The enhancement of the low-energy excitations is a common characteristic among the different defects deposited from the Au STM-tip. This behavior is illustrated in Figure 4B for another defect. The inset of this figure also shows difference spectra for several different defects with similar behavior (see caption for details). Another common characteristic is that the impurity-induced excitations are localized to within lateral distances of about 20\AA\ around the impurities. This spatial characteristic was also imaged by measuring the AC $dI/dV$ at a fixed voltage while scanning the tip in constant DC current mode. Such an image is shown in Figure 3B, along with the constant current topographs (Figure 3A) acquired simultaneously from an area of the surface which includes two defect structures. The bound excitations for each of the defects are localized at the dark regions in this gray scale image.\cite{Details}

The data above demonstrate the central result of this paper, that non-magnetic defects alter the local LDOS of Bi$_{2}$Sr$_{2}$CaCu$_{2}$O$_{8}$ surface by inducing low-energy excitations. These excitations are bound to the impurities on length scales comparable to the superconducting coherence length. We emphasize that similar experiments on a Nb surface have previously shown the LDOS of a conventional superconductor to be insensitive to the presence of atomic-scale non-magnetic defects deposited from a Au tip \cite{Yazdani}.

Theoretical works on impurity scattering in a d-wave superconductor have predicted that an isolated non-magnetic defect gives rise to a semi-bound excitation with energy $E_{B}<\Delta_{max}$. \cite{Balatsky1} In this theoretical picture, $E_B$ is determined by the strength of the impurity scattering potential, with $E_{B}=E_{F}$ for scattering in the unitarity limit. More recent efforts show that including the local suppression of the superconducting order parameter in the calculations drives the effective scattering strength of even a strong impurity to the unitarity limit.\cite{Shnirman} The impurities discussed above modify the LDOS over a range of energies, many of which induce sharp resonances very close to $E_{F}$, and are in the unitarity limit. The asymmetric or splitting of measured resonances near $E_{F}$ support the possibility that impurities locally break the electron-hole symmetry \cite{Balatsky1}. However, the underlying electronic states of Bi$_{2}$Sr$_{2}$CaCu$_{2}$O$_{8}$ has also been suggested as a cause for the asymmetry.\cite{Flatte98}. 

The spatial structure of the resonance for an isolated impurity has been predicted to be cross-shaped, reflecting the anisotropy in the superconducting order parameter.\cite{Balatsky1} Theoretical efforts have emphasized that the spatial and angular character of single-impurity resonances greatly affect their overlap and the nature of the many-impurity state. The overlap between these resonances for bulk impurities determines the formation of low-energy impurity bands and the behavior of the thermodynamic properties at low temperatures.\cite{Balatsky1,Balatsky2}.  The single impurity-induced excitations measured here appear to be localized to a region of about a coherence length, and do not show the predicted cross-shaped pattern, which has a delocalized nature along the nodes of the d-wave order parameter. More detailed high-resolution spatial measurements may be required to resolve this fine structure.

The observation of the impurity-induced excitation is also intriguing within the context of
local electronic properties of a d-wave superconductor near other spatial perturbations such as surfaces, crystal twins, and grain boundaries. The scattering from these boundaries is expected to give rise to Andreev bound states whenever the incident and reflected quasi-particles experience the sign change of the d-wave order parameter, i.e. when the boundary is normal to the direction in momentum space along d-wave nodes.\cite{Hu,Walker2,Sigrist} Directional tunneling spectroscopy in planar tunnel junction\cite{Green}, point contact spectroscopy\cite{Wei}, and grain boundary tunnel junctions\cite{Alff} have shown evidence for such Andreev states. The experimental data reported here show that the scattering at isolated atomic-scale impurities can also induce similar resonances at $E_F$. In fact, a theoretical connection between scattering processes at impurities and those at interfaces can be made within a semi-classical approximation.\cite{Chen,Innac} Perturbations such as defects or surfaces cause scattering and interference between electronic states from different regions of k-space. These processes together with the sign change of the order parameter inherent to a d-wave superconductor give rise to the zero energy resonances observed in the experiments.

In conclusion, we have shown that ordinary non-magnetic impurities can induce localized low-energy excitations in a d-wave superconductor. For some of the impurities, we observed the impurity-induced resonance to be essentially in the middle of the gap at $E_{F}$, as predicted for impurity scattering in the unitarity limit. Such a zero bias state is also the hallmark of Andreev scattering in a d-wave superconductor whenever scattering of the quasi-particles can explore the sign change of the d-wave order parameter. The experimental data described here were obtained at IBM Almaden Research Center. It is our pleasure to thank  L. H. Greene, D. J. Van Harlingen, J. A. Sauls, G. Blumberg, D. K. Campbell, and especially A. V. Balatsky, M. E. Flatt\'{e}, A. Shnirman, I. Adagideli, and  P. M. Goldbart for fruitful discussions.

{\it Note Added:} After submission of this manuscript, we have become aware of another STM study of defects in Bi$_{2}$Sr$_{2}$CaCu$_{2}$O$_{8}$.\cite{Hudson}.


\begin{references}
\bibitem[*]{byline} e-mail ayazdani@uiuc.edu.

\bibitem{Vanharlingen}D.J. Van Harlingen, Rev. Mod. Phys. {\bf 67}, 515 (1995).
\bibitem{Annette} For a review see, J.F. Annett, N. Goldenfeld, and A.J. Leggett in {\it Physical Properties of High Temperature Superconductors V}, edited by D.M. Ginsberg, World Scientific, p. 375 (1996).
\bibitem{Lee}P.A. Lee, Phys. Rev. Lett. 71, 12, 1887-1890 (1993); P.J. Hirschfeld and N. Goldenfeld, Phys. Rev. B {\bf 48}, 4219 (1993).
\bibitem{Byers}J.M. Byers, M.E. Flatt\'{e}, and D.J. Scalapino, Phys. Rev. Lett. {\bf 71}, 3363 (1993).
\bibitem{Balatsky1}A.V. Balatsky, M.I. Salkola, and A. Rosengren, Phys. Rev. B {\bf 51}, 15547 (1995); 
M.I. Salkola, A. V. Balatsky and D.J. Scalapino , Phys. Rev. Lett. {\bf 77}, 184 (1996).
\bibitem{Balatsky2}A.V. Balatsky, A. Rosengren, and B. L. Altshuler, Phys. Rev. Lett. {\bf 73}, 720 (1994); A.V. Balatsky and M.I. Salkola, Phys. Rev. Lett. {\bf 76}, 2386, (1996).
\bibitem{Franz1}M. Franz, C. Kallin, and A.J. Berlinsky, Phys. Rev. B {\bf 54}, R6897 (1996).
\bibitem{Walker1}M.E. Zhitomirsky and M.B. Walker, Phys. Rev. Lett. {\bf 80}, 5413 (1998).
\bibitem{Buchholtz} L.J. Buchholtz and G. Zwichnagl, Phys. Rec. B {\bf 23}, (1981). 
\bibitem{Hu}C.-R. Hu, Phys. Rev. Lett. {\bf 72}, 1526 (1994); Y. Tanaka and S. Kashiwaya, Phy. Rev. Lett. {\bf 74}, 3451 (1995).
\bibitem{Green}M. Covington {\it et al.}, Phys. Rev. Lett. {\bf 79}, 277 (1997).
\bibitem{Sauls}M. Fogelstr\"{o}m, D. Rainer, and J.A. Sauls, Phys. Rev. Lett.{\bf 79}, 281 (1997).
\bibitem{Walker2}M. E. Zhitomirsky and M. B. Walker, Phys. Rev. Lett. {\bf 79}, 1734 (1997). 
\bibitem{Sigrist}W. Belzig, C. Burder, and M. Sigrist, Phys. Rev. Lett. {\bf 80}, 4285 (1997).
\bibitem{Yazdani}A. Yazdani, B.A. Jones, C.P. Lutz, M.F. Crommie, and D.M. Eigler, Science {\bf 275}, 1767 (1997).
\bibitem{Shnirman}A. Shnirman, I. Adagideli, P. M. Goldbart, and A. Yazdani, cond-mat/9903252, (1999).
\bibitem{Wei} J.Y.T. Wei, N.-C. Yeh, D.F. Garrigus, and M. Strasik, Phys. Rev. Lett. {\bf 81}, 2542 (1998).
\bibitem{Alff}L. Alff {\it et al.}, Phys. Rev. B {\bf 58}, 11197 (1998). 
\bibitem{Kirk} M.D. Kirk {\it et al.}, Appl. Phys. Lett. {\bf 52}, 2071 (1988), see also S.H. Pan, E.W. Hudson, J. Ma, and J.C. Davis, Appl. Phys. Lett. {\bf 73}, 58 (1998).
\bibitem{Contact}J. K. Gimzewski and R. M\"{o}ller, Phys. Rev. B {\bf 36}, 1284 (1987).
\bibitem{Renner}Ch. Renner {\it et al.}, Phys. Rev. Lett. {\bf 80}, 3606 (1998). In particular, see the data on overdoped sample with T$_c$=74.3K.
\bibitem{Details} The reverse contrast comes about because we choose a DC bias voltage corresponding to energies larger than the impurity-induced excitations, where the bound excitation affects $dI/dV$ only indirectly by contributing to the total current, $I$, so that the tip withdraws and reduces $dI/dV$.
\bibitem{Flatte98}M.E. Flatt\'{e} and J.M. Byers, Phys. Rev. Lett. {\bf 80}, 4546 (1998).
\bibitem{Chen} See also, D.C. Chen, D. Rainer, J.A. Sauls, Proceedings of the Verditz Workshop 
on Quasiclassical Methods in Superconductivity, Afritz, Austria (1996).
\bibitem{Innac}I. Adagideli, A. Shnirman, P.M. Goldbart, and A. Yazdani, in preparation.
\bibitem{Hudson} E. Hudson, S. H. Pan, A. K. Gupta, K-W Ng, and J. C. Davis to appear in Science.

\end{references}
\end{document}